\documentstyle[aps,preprint]{revtex}
\begin{document}
\title{Field Theories for Learning Probability Distributions}
\author{William Bialek,$^1$ Curtis G. Callan,$^2$ and Steven P. Strong$^1$}

\address{$^1$NEC Research Institute,
4 Independence Way,
Princeton, New Jersey 08540\\
$^2$Department of Physics,
Princeton University,
Princeton, New Jersey 08544}

\date{\today}

\maketitle

\begin{abstract}
Imagine being shown $N$ samples of random variables drawn
independently from the same distribution.  What can you say about the
distribution?  In general, of course, the answer is
nothing, unless you have some prior notions about what to expect.
From a Bayesian point of view one needs an {\it a priori}
distribution on the space of possible probability distributions,
which defines a scalar field theory.
In one dimension, free field theory with a normalization constraint
provides a tractable formulation of the
problem, and we discuss generalizations to higher dimensions.
\end{abstract}
\vfill\newpage


As we watch the successive flips of a coin (or the meanderings of stock prices),
we ask ourselves if what we see is consistent with the conventional
probabilistic model of a fair coin.  More quantitatively, we might try to fit
the data with a definite model that, as we vary parameters, includes the fair
coin and a range of possible biases.  The estimation of these underlying
parameters is the classical problem of statistical inference or `inverse
probability,' and has its origins in the foundations of probability
theory itself \cite{history}.
But when we observe continuous variables, the relevant
probability distributions are functions, not finite lists of numbers as in the
classical examples of flipping coins or rolling dice.  In what sense can we
infer these functions from a finite set of examples?  In particular, how do we
avoid the solipsistic inference in which each data point we have observed is
interpreted as the location of a narrow peak in the underlying distribution?

Let the variable of interest be $x$ with probability distribution
$Q({x})$; we start with the
one dimensional case.  We are given a set of points
$x_1 , x_2 , \cdots ,  x_N$ that are drawn independently from $Q( x)$, and are
asked to estimate $Q( x)$ itself.  One approach is to assume that 
all possible $Q(x)$ are drawn from a space parameterized by
a finite set of coordinates,
implicitly excluding distributions
that have many sharp features.  In this case, it is
clear that the number of examples $N$ can eventually overwhelm the number of
parameters $K$ \cite{finite}.   Although the finite dimensional case is often
of practical interest, one would like a formulation faithful to the
original problem of estimating  a function rather than a limited number of
parameters.

No finite number of examples will  determine uniquely the whole
function $Q(x)$, so we require a probabilistic description.
Using Bayes' rule, we can write the
probability of the function $Q(x)$ given the data as
\begin{eqnarray}
&&P[Q(x)| x_1 , x_2 , ... , x_N]
\nonumber
\\
&&\,\,\,\,\,
 = 
{{P[x_1 , x_2 , ... , x_N | Q(x)] P[Q(x) ]}
\over {P(x_1 , x_2 , ... , x_N )}} \\
&&\,\,\,\,\, =
{{Q(x_1 ) Q(x_2 ) \cdots Q(x_N) P[Q(x)]}
\over{\int [dQ(x)] Q(x_1 ) Q(x_2 ) \cdots Q(x_N) P[Q( x)]}} ,
\label{conditional}
\end{eqnarray}
where we make use of the fact that each $x_i$ is chosen
independently from the distribution $Q(x)$, and $P[Q(x) ]$
summarizes our {\it a priori} hypotheses about the form of $Q(x)$. 
If asked for an explicit estimate of $Q(x)$,
one might try to optimize the estimate so that
the mean square deviation from the correct answer is, at each point $x$,
as small as possible.  This optimal least-square estimator
$Q_{\rm est} (x; \{x_i\} )$ is the average of $Q(x)$ in
the conditional distribution of Eq. (\ref{conditional}),
which can be written as
\begin{equation}
Q_{\rm est} (x; \{x_i\} )
=
{{\langle Q(x) Q(x_1 ) Q(x_2 ) \cdots Q(x_N) \rangle ^{(0)}}\over
{\langle Q(x_1 ) Q(x_2 ) \cdots Q(x_N) \rangle^{(0)}}} ,
\label{est}
\end{equation}
where by $\langle \cdots \rangle^{(0)}$ we mean expectation values with
respect to the {\it a priori} distribution $P[Q(x)]$.   The prior distribution
$P[Q(x)]$ is  a scalar field theory, and the n-point functions of this theory
are precisely the objects that determine our inferences from the data.

The restriction of the distribution $Q(x)$ to a finite dimensional space
represents, in the field theoretic language, a sharp ultraviolet cutoff
scheme.  Several authors have considered the problem
of choosing among distributions with different numbers of parameters,
which corresponds to assuming that the true theory, $P[Q(x)]$, has a
hard ultraviolet cutoff whose unknown location is to be 
set by this choice.
As in field theory itself, one would like to have a
theory in which one can remove the cutoff
without any unpleasant consequences. Our Bayesian
approach will provide this.

The prior distribution, $P[Q(x)]$, should
capture our prejudice that the distribution
$Q(x)$ is smooth, so $P[Q(x)]$ must
penalize large gradients, as in conventional field
theories.  To have a field variable $\phi (x)$ that takes on a full range of
real values $(-\infty < \phi < \infty)$, we write
\begin{equation}
Q(x) = {1\over\ell} \exp[-\phi(x)] ,
\end{equation}
where $\ell$ is an arbitrary length scale.
Then we take $\phi$ to be a free scalar field with a  constraint to enforce
normalization of $Q(x)$.  Thus $\phi(x)$ is chosen from a probability
distribution
\begin{eqnarray}
P_\ell[\phi(x)]
& = & {1\over Z} \exp\left[ -{\ell\over2} \int dx (\partial_x \phi)^2
\right]\nonumber\\
&&\,\,\,\,\,\times
\delta \left[ 1 - {1\over \ell} \int dx e^{-\phi(x)}\right] ,
\end{eqnarray}
where we write $P_\ell [\phi(x)]$ to remind us that
we have chosen a particular value for $\ell$;
we will later consider averaging over a distribution 
of $\ell$'s, $P(\ell )$.
The objects of interest are the correlation functions:
\begin{eqnarray}
&&\langle Q(x_1 ) Q(x_2 ) \cdots Q(x_N) \rangle^{(0)}
\nonumber\\
&&\,\,\,\,\,
=
\int D\phi P[\phi(x) ] \prod_{i=1}^N {1\over\ell} \exp[-\phi(x_i)] \\
&&\,\,\,\,\, =
{1\over{\ell^N}} {1\over Z} \int {{d\lambda }\over{2\pi}}
\int D\phi
\exp\left[ - S(\phi ; \lambda ) \right] ,
\label{problem}
\end{eqnarray}
where, by introducing the Fourier representation of the
delta function,  we define the action
\begin{eqnarray}
S(\phi ; \lambda ) &=&
{\ell\over2} \int dx (\partial_x \phi)^2 \nonumber\\
&&\,\,\,\,\,
+i{\lambda\over\ell} \int dx e^{-\phi(x)}
+\sum_{i=1}^N \phi (x_i ) - i\lambda
.
\end{eqnarray}
We evaluate the functional integral in Eq. (\ref{problem}) in a 
semiclassical approximation, which becomes accurate as $N$ becomes
large.   Keeping only the configuration corresponding to extremizing
the action---the pure classical approximation, with no fluctuations---
is equivalent to maximum likelihood estimation,  which
chooses the distribution, $Q(x)$, that maximizes $P[Q(x) |\{x_i\}]$.
In our case, integration over fluctuations will play a crucial
role in setting the proper value of the scale $\ell$.

The classical equations of motion for $\phi$ and $\lambda$ are, as usual,
\begin{equation}
{{\delta S(\phi ; \lambda )}\over {\delta \phi(x)}} = 
{{\partial S(\phi ; \lambda )}\over {\partial \lambda}} = 0,
\end{equation}
which imply
\begin{eqnarray}
\ell \partial_x^2 \phi_{\rm cl}(x) +i{\lambda_{\rm cl}\over\ell}
e^{-\phi_{\rm cl}(x)} &=& \sum_{i=1}^N \delta (x-x_i)
\label{phieqn}
\\
{1\over\ell} \int dx e^{-\phi_{\rm cl}(x)} &=& 1 .
\label{lambdaeqn}
\end{eqnarray}
Integrating Eq. (\ref{phieqn}) and comparing with Eq. (\ref{lambdaeqn}), we
find that $i\lambda_{\rm cl} = N$, provided that $\partial\phi (x)$
vanishes as $|x| \rightarrow \infty$ \cite{explicit}.  If the points
$\{x_i\}$ are actually chosen from a distribution $P(x)$, then, as
$N\rightarrow\infty$, we hope that  $\phi_{\rm cl} (x)$ will converge to
$-\ln[\ell P(x)]$.  This would guarantee that our average over all possible
distributions
$Q(x)$ is dominated by configurations $Q_{\rm cl}(x)$ that approximate the true
distribution.
So we write $\phi_{\rm cl}(x) = -\ln[\ell P(x)] + \psi(x)$ and
expand  Eq. (\ref{phieqn})
to first order in $\psi (x)$.  In addition we notice that the sum of
delta functions can be written as
\begin{equation}
\sum_{i=1}^N \delta (x-x_i) = N P(x) + \sqrt N  \rho(x) ,
\end{equation}
where $\rho(x)$ is a fluctuating density such that
\begin{equation}
\langle \rho (x) \rho (x')\rangle = P(x)\delta(x-x').
\end{equation}
The (hopefully) small field $\psi(x)$ obeys the equation
\begin{equation}
\left[\ell\partial_x^2  - NP(x)\right]\psi(x)
= \sqrt N \rho (x) + \ell\partial_x^2\ln P(x),
\end{equation}
which we can solve by WKB methods because of the large factor $N$:
\begin{eqnarray}
\psi(x) &=&
\int dx' K(x , x' ) \left[ \sqrt N \rho (x') + \ell\partial_x^2\ln P(x') \right]
\\
K(x, x') &\sim&
{1\over {2\sqrt N}}
\left[ \ell^2 P(x)P(x')\right]^{-1/4}
\nonumber\\
&&\,\,\,\,\,\times
\exp\left[ - \int_{\min (x,x')}^{\max (x,x')} dy \sqrt{{NP(y)}\over\ell}
\right] .
\end{eqnarray}
Thus the ``errors'' $\psi(x)$ in our estimate of the distribution
involve an average of the fluctuating density over a region of (local) size
$\xi \sim [\ell/NP(x)]^{1/2}$.  The average systematic error and the
mean-square random error are easily computed in the limit $N\rightarrow\infty$
because this length scale becomes small.
We find
\begin{eqnarray}
\langle \psi (x) \rangle
&=&
{\ell\over{NP(x)}} \partial_x^2 \ln P(x)  + \cdots ,
\label{sys}
\\
\langle [\delta\psi(x)]^2\rangle 
&=&
{1\over 4} {1\over\sqrt{NP(x)\ell}}  + \cdots ,
\label{random}
\end{eqnarray}
Higher moments also decline as powers of $N$,
justifying our claim that 
the classical solution converges to the correct distribution.

The complete semiclassical form of the n-point function is
\begin{eqnarray}
&&
\langle Q(x_1 ) Q(x_2 ) \cdots Q(x_N )\rangle^{(0)}
\nonumber\\ && \,\,\,\,\,
\approx {1\over {\ell^N}} R \exp\left[ -S(\phi_{\rm cl};\lambda =-iN )\right] ,
\end{eqnarray}
where $R$ is  the ratio of determinants,
\begin{equation}
R = \left[
{{\det(-\ell\partial_x^2 + N Q_{\rm cl} (x))}
\over
{\det(-\ell\partial_x^2 )}}
\right]^{-1/2} .
\end{equation}
This has to be computed a bit carefully---there is no restoring force for
fluctuations $\lambda$, but these can be removed by fixing the 
spatially uniform component of $\phi(x)$, which 
enforces normalization of $Q(x)$.   Since everything is finite in
the infrared this is does not pose a  problem \cite{corrections}.
Then the computation of the
determinants is standard \cite{coleman}, and we find
\begin{equation}
R = \exp\left[- {1\over 2}\left({N\over{\ell}}\right)^{1/2}
\int dx \sqrt{Q_{\rm cl} (x)} \right],
\end{equation}
where as before we use the limit $N\rightarrow\infty$ to simplify
the result \cite{worry}.  It is interesting to note
that $R$ can also be written as $\exp[-(1/2)\int dx ~\xi^{-1}]$,
so the fluctuation contribution to the effective action
counts the number of independent ``bins'' 
(of size $\sim\xi$)  that are
used in describing the function $Q(x)$.

Putting the factors together, we find that
\begin{eqnarray}
&&\langle Q(x_1 ) Q(x_2 ) \cdots Q(x_N )\rangle^{(0)}
\nonumber\\
&&\,\,\,\,\,
\approx \prod_{i=1}^N P(x_i) \exp [-F(x_1 , x_2 ,  \cdots , x_N )] ,
\end{eqnarray}
where the correction term $F$ is given by
\begin{eqnarray}
F(\{x_i\}) &=& {1\over 2}\left( {N\over{\ell}}\right)^{1/2} \int dx P^{1/2} (x)
e^{-\psi(x)/2}
\nonumber\\
&&\,\,\,\,\,
+ {\ell\over 2} \int dx (\partial_x \ln P - \partial_x \psi)^2
+ \sum_{i=1}^N \psi(x_i) 
\end{eqnarray}
One might worry that $\psi(x)$ is driven by density fluctuations that
include delta functions at points $x_i$, while, when we evaluate $F$,
we sum up
the values of $\psi(x)$ precisely at these singular points.  In fact,
these 
terms are finite and of the same order of magnitude as the
fluctuation determinant.
Similarly, our estimate of the probability
distribution from Eq. (\ref{est}) is finite even when we ask about $Q(x)$ at
the points where we have been given examples.  This is not so surprising---we
are in one dimension where ultraviolet divergences should not be a problem.

Although our theory is finite in the ultraviolet, we do have an arbitrary
length scale $\ell$.  This means that we define,  {\it a priori}, a scale on
which variations of the probability $Q(x)$ are viewed as ``too fast.''
One would rather  let all scales in our estimate of the
distribution $Q(x)$ emerge from the data points themselves.
We can restore scale invariance (perhaps scale indifference is a
better term here) by viewing $\ell$ itself as a parameter that needs to be
determined.  Thus, as a last step in evaluating the functional integral, we
should integrate over $\ell$, weighted by some prior distribution,
$P(\ell)$, for
values of this parameter.  The hope is that this integral will be dominated
by some scale, $\ell_*$, that is determined primarily by the structure of $Q(x)$
itself, at least in the large $N$ limit.
As long as our {\it a priori}
knowledge about $\ell$ can be summarized by a reasonably smooth
distribution, then, at large $N$,  $\ell_*$ must be the
minimum of $F$, since this is the only place where $\ell$ appears
with coefficients that grow as powers of $N$.
To see how this  works we compute the average value of $F$ and 
minimize with respect to $\ell$. The result is
\begin{equation}
\ell_* = N^{1/3}
\left[
{{(5/8)\int dx P^{1/2}}\over{\int dx (\partial_x \ln P)^2}}
\right]^{2/3} .
\label{ell}
\end{equation}
Strictly, one should use a particular value
of $F$ and not its average, but fluctuations are of lower order
in $N$ and do not change the qualitative result $\ell_* \propto
N^{1/3}$.

The semiclassical evaluation of the relevant functional integrals
thus gives a classical configuration that smooths the examples on a scale
$\xi \propto (\ell/N)^{1/2}$, and the scale $\ell$ is selected by a
competition between the classical kinetic energy or smoothness
constraint and the fluctuation determinant.  If the fluctuation effects
were ignored, as in maximum likelihood estimation,
$\ell$ would be driven to zero and we would be overly
sensitive to the details of the data points.
This parallels the discussion of ``Occam factors'' in the finite
dimensional case, where the phase space factors from integration over
the parameters $\{g_\mu\}$ serve to discriminate against models with
larger numbers of parameters \cite{finite}.
It is not clear from the discussion of finite dimensional models,
however, whether these factors are sufficiently powerful to reject
models with an infinite number of parameters.
Here we see that, even in an infinite dimensional setting,
the fluctuation terms are sufficient to control the estimation
problem and select a model with finite, $N$-dependent, complexity.

Because we are trying to estimate a function, rather
than a finite number of parameters, we must allow ourselves to give a
more and more detailed description of the function $Q(x)$ as we see more
examples; this is quantified by the scale
$\xi_*$ on which the estimated distribution is forced to be smooth.
With the selection of the optimal
$\ell$ from Eq. (\ref{ell}),  we see that
$\xi_* \propto N^{-1/3}$.
The classical solution converges
to the correct answer with a systematic error, from Eq. (\ref{sys}),
that vanishes as $\langle \psi \rangle \propto N^{-2/3}$,
while the random errors have a variance [Eq. (\ref{random})]
that vanishes  with the same power of $N$.
We can understand this result by noting that in a region of
size $\xi_*$ there are, on average, $N_{\rm ex} \sim N P(x) \xi_*$
examples, which scales as $N_{\rm ex} \propto N^{2/3}$;
the random errors then have a standard deviation $\delta\psi_{\rm rms}
\sim 1/\sqrt{N_{\rm ex}}$ \cite{connections}.

How does this discussion generalize to higher dimensions?
If we keep the simple free field theory then we will
have problems with ultraviolet divergences in the various correlation
functions of the field $\phi (x)$.  Because 
$Q(x) = (1/\ell ) \exp[-\phi(x)]$, ultraviolet divergences
in $\phi$ mean that we cannot define a normalizable distribution for
the possible values of $Q$ at a single point in the continuum limit.
In terms of information theory \cite{shannon},
if functions $Q(x)$ are drawn from 
a distribution functional with ultraviolet divergences, then even
specifying the function $Q(x)$ to finite precision requires an
infinite amount of information.

As an alternative, we can consider higher derivative actions
in higher dimensions.  All the calculations are analogous to those
summarized above, so here we only list the results.  If we write,
in $D$ dimensions, $Q(x) = (1/\ell^D)\exp[-\phi(x)]$,
and choose a prior distribution
\begin{eqnarray}
P[\phi(x)] & = & {1\over Z} \exp\left[ -{{\ell^{2\alpha -D}}\over2}
\int d^D x (\partial^\alpha_x
\phi)^2\right]\nonumber\\
&&\,\,\,\,\,\times
\delta \left[ 1 - {1\over {\ell^D}} \int d^Dx e^{-\phi(x)}\right] ,
\label{multidim}
\end{eqnarray}
then to insure finiteness in the ultraviolet we must have
$2\alpha > D$.  The saddle point equations lead to a distribution
that smooths the examples on a scale $\xi \sim  (\ell^{2\alpha -D}
/NQ)^{1/2\alpha}$, and fluctuation determinant makes a contribution to
the action $\propto \int d^Dx [NQ(x)/\ell^{2\alpha-D}]^{D/2\alpha}$.
Again we find the optimal value of $\ell$ as
a compromise between this term and the kinetic energy, resulting in
$\ell_* \propto N^{D/(4\alpha^2 -D^2)}$.
Then the optimal value of $\xi$ becomes
$\xi_* \propto N^{-1/(2\alpha+D)}$,
so that the estimated distribution is smooth
in volumes of dimension $\xi_*^D$ that contain
$N_{\rm ex} \sim NQ\xi^D \sim N^{2\alpha/(2\alpha +D)}$
examples. Then the statistical errors in the estimate will
behave as
\begin{equation}
\delta\psi_{\rm rms} \propto \delta Q / Q \sim N_{\rm ex}^{-1/2} \sim N^{-\mu}
,\end{equation}
with the ``error exponent''
$\mu =\alpha/(2\alpha +D)$.
Note that since $2\alpha >D$, the exponent $1/4 < \mu < 1/2$.
The most rapid convergence, $\mu = 1/2$,
occurs if $Q(x)$ is drawn from a family
of arbitrarily smooth ($\alpha \rightarrow\infty$) distributions,
so we can choose fixed, small bins in which to accumulate the samples,
leading to the naive $1/\sqrt N$  counting statistics. If we assume that our
prior distribution functional is local, then $\alpha$ must be an integer
and we can have $\mu \rightarrow 1/4$ only as $D\rightarrow\infty$,
so that the slowest possible convergence occurs in infinite dimension.

The fact that higher dimensional functions are more difficult to learn
is often called the `curse of dimensionality.'  
We see that this is not just a quantitative problem---unless we
hypothesize that higher dimensional functions are drawn from ensembles
with proportionately higher order notions of smoothness, one would
require an infinite amount of information to specify the function at
finite precision.  Once we adopt these more stringent smoothness
hypotheses, however, the worst that happens is a reduction in
the error exponent $\mu$ by a factor of two.

Is there a more general motivation for the choice of action in Eq.
(\ref{multidim})?  First, we note that this action gives the maximum
entropy distribution consistent with a fixed value of $\int d^D x
(\partial_x^\alpha \phi)^2$, and by integrating over $\ell$ we integrate
over these fixed values. Thus our action is equivalent 
to the rather generic assumption that our probability
distributions are drawn from an ensemble 
in which this ``kinetic energy'' is finite.
Second, addition of a constant to $\phi (x)$ can be absorbed in a
redefinition of $\ell$, and since we integrate over $\ell$
it makes sense to insist on $\phi \rightarrow \phi + {\rm const.}$ as a
symmetry.
Finally, addition of other terms to the action cannot change the 
asymptotic behavior at large $N$ unless these terms
are relevant operators in the ultraviolet.
Thus many different priors $P[Q(x)]$ will exhibit the same asymptotic
convergence properties, indexed by a single exponent $\mu (\alpha )$.

We thank V. Balasubramanian, R. Koberle, S. Omohundro, and W. Smith for
helpful discussions and for comments on the manuscript.


\begin{thebibliography}{99}
\bibitem{explicit}
Direct integration of Eq. (\ref{phieqn}) shows that, for $x$ outside
the region sampled by the $\{x_i\}$, 
$Q_{\rm cl} (|x| > {\rm max}|x_i|) = 2\ell/Nx^2$,
so that $\partial\phi(x)$ does  vanish
at infinity.
Note also that the
total weight assigned to the unobserved region vanishes as
$N$ increases.  This explicit integration can be continued
into the sampled region, reducing the problem
to quadratures.  At the end,
however, one has to solve the matching conditions at each point
$x_i$, and this leads to complicated expressions.  Since we are
interested in showing that the solution converges to the correct
answer, we find it more efficient to study perturbation theory around
this answer and see if it is consistent.
\bibitem{corrections}
We keep track of terms that contribute positive powers of $N$
in the effective action; the zero mode that we need to exclude gives at
most $\ln N$.
\bibitem{worry}
With the asymptotic behavior of $Q_{\rm cl} (x)$ noted
in \cite{explicit}, the WKB integral does not converge.
More precisely the WKB approximation breaks down far
outside the region where we have observed the samples $\{ x_i\}$.
But the total weight in this region declines with $N$,
so proper treatment of these terms generates subleading
terms in the effective action.
\bibitem{connections}
This scaling behavior is also found by
Barron and Sheu in Ref. \cite{finite}.
They studied the learning of probability distributions
for which $\phi (x)$ lives
in an  m-dimensional linear space, and then they choose m to optimize
convergence.  The exponents derived here correspond to the case
where $\phi (x)$ is continuous and has a first derivative 
that is square integrable, which has a clear connection to our
choice of prior.  W. Smith has emphasized to us that this scaling
can be derived most simply for piecewise constant approximations
to $Q(x)$---conventional binning procedures,
with the bin size chosen to minimize the sum of systematic
and statistical errors.  In the field theoretic language binning
is a lattice cutoff.  We find it attractive that these
more rigorous mathematical results can be recovered in a Bayesian
formulation that has no explicit cutoff.
\end{thebibliography}
\end{document}